# GENETIC CODE AS A HARMONIC SYSTEM: THREE SUPPLEMENTS

Miloje M. Rakočević

*Department of Chemistry, Faculty of Science, University of Niš, Višegradska 33, Serbia (E-mail: mirkovmile@mts.rs)(www.rakocevcode.rs)*

**Abstract**.

The paper represents three supplements to the source paper, q-bio/0610044 [q-bio.OT], with three new series of harmonic structures of the genetic code, determined by Gauss arithmetical algorithm; by Table of Minimal Adding, as in (Rakocevic, 2011a: Table 4; 2011b: Table 4); all structures in relation to Binary-code tree (Rakocevic, 1998). The determination itself is realized through atom and nucleon number balancing and nuancing of molekular polarity. In first supplement the word is about some additional harmonic structures in relation to a previous our paper (Rakocevic, 2004); in the second one about the relation that structures with the polarity of protein amino acids. In the third supplement we give new ideas about the genetic code by an inclusion of the notions cipher of the genetic code and the key of that cipher.

## SUPPLEMENT 1. *Some additional harmonic structures*

**1. Introduction.** In this supplement we give a new series of harmonic structures of the genetic code, determined by Gauss arithmetical algorithm as it is shown in the source paper (Rakočević, 2006a), q-bio.OT/0610044. In source paper beside others, we showed that 16 non-contact canonical amino acids (AAs), in an amino acid molecule size order, are determined by a Gauss' arithmetical algorithm. In this supplement however we show that the same is valid – for the same 16 non-contact amino acids – for their coding order (ordinal number) in Genetic Code Table (GCT). The difference is in the fact that here are valid only two last steps from the said Gauss' algorithm. Namely, from all Gauss' algorithm quantums – 11, 21, 31, 41,( ), 61, 71, 81, 91 – only two last steps are here in the „game"; the quantums „81" and „91" as well as their arithmetical mean 86±0. (Cf. source paper-Figure 1 and see the positions of quantums 86±1 in source paper-Figure 1.1.) As in source paper, all determinations are realized through principle of minimum change, i.e. through the unit atom number balances in first or in the second position of the digitnumber-notation, respectively (x±00, x±01; x±10), (y±00, y±01; y±10).

**2. Results and Discussion.** In following 16 illustrations (Tables) are given the results of calculations of atom number within amino acid side chains; the calculations, related to the amino acid system built either from only 16 non-contact or from all 20 AAs (16 non-contact plus 4 contact AAs). By this the rest of four contact AAs make: Glycine (G), Proline (P), Valine (V) and Isoleucine (I). After our hypothesis (Hypothesis 1) there are some other possibilities of the amino acid splitting into 4 special and 16 other AAs; such a splitting which is related to the atom number balances (x±00, x±01; x±10), (y±00, y±01; y±10). The possible examples are: Serine (S), Threonine (T), Cisteine (C) and Methionine (M) as chalcogene AAs (chalcogene because they possess oxygen or sulfur in molecule side chains); then carboxylic AAs and their amide derivatives: Aspartic acid (D), Glutamic acid (E), Asparagine (N) and Glutamine (Q); four aromatic versus 16 aliphatic etc.

| | | | | | |
|---|---|---|---|---|---|
| 01 | F | 14 | 11 | Q | 09 |
| 02 | L | 13 | 08 | N | 10 |
| 03 | M | 11 | 15 | K | 11 |
| 04 | S | 05 | 07 | D | 12 |
| 05 | T | 08 | 10 | E | 13 |
| 06 | A | 04 | 05 | C | 14 |
| 07 | Y | 15 | 18 | W | 15 |
| 08 | H | 11 | 17 | R | 16 |
| Odd | | 48 | 54 → | 102± 0 | |
| Even | | 33 | 37→ | 102-32 | |
| | | 81 | 91 | | |
| | | ______ | 86-1 | | |
| | | - - - - - | 86+1 | | |
| Odd | | 16 | 48 (68) | | |
| Even | | 20 | 52 (68) | | |
| 01 | F | 14 | 17 | R | 16 |
| 02 | L | 13 | 18 | W | 15 |
| 03 | M | 11 | 05 | C | 14 |
| 04 | S | 05 | 10 | E | 13 |
| 05 | T | 08 | 07 | D | 12 |
| 06 | A | 04 | 15 | K | 11 |
| 07 | Y | 15 | 08 | N | 10 |
| 08 | H | 11 | 11 | Q | 09 |
| Odd | | 48 | 37→ | 86-1 | |
| Even | | 33 | 54→ | 86+1 | |
| | | 81 | 91 | | |
| | | ______ | 102 ±0 | | |
| | | - - - - - | 102-32 | | |
| Odd | | 16 | ----52 (68) | | |
| Even | | 20 | ----48 (68) | | |

**Table 1 [left].** This Table is analog with Table 1.2 in source paper. The 16 non-contact AAs arranged into two octets, correspondingly to their ordinal number in GCT, i.e. correspondingly to ordinal number of belonging codons; the first octet on the left and the second one (in the sequence up/down) on the right. The calculation for the ordinal numbers for all 16 amino acids (their sums) is given just down (16+20 = 36 and 48 + 52 = 100).

**Table 2 [right].** All is the same as in previous Table, except the second octet (on the right) is given in a vice-versa sequence (down/up).

*Remark* 1: For the terms „contact“ and „non-contact“ AAs as well as for all other new terms see source paper q-bio.OT/0610044.

*Remark* 2: Hypothesis 1 follows from the idea that quantum „4“ represents the first possible case for the existence of *one pair of pairs* (cf. legend given for Figure 1 in source paper). On the other hand the equation 16 + 4 = 20, corresponding to square equation $x^2 + x = 20$, is one and only case from the family of metalic means, in relation to the Golden mean (Rakočević, 1998, 2004a; 2006b – Table 2), which corresponds with the middle point in the Harmonic multiplication table.

The first two Tables (Tables 1 & 2) show a correspondence with the natural numbers series from one side and with the Gauss’ arithmetical algorithm from the other side (with quantums “81” and “91” and their arithmetic mean 86±0 and/or 86±1).

| | | | | | |
|---|---|---|---|---|---|
| 06 | A | 04 | 05 | C | 14 |
| 04 | S | 05 | 07 | D | 12 |
| 05 | T | 08 | 10 | E | 13 |
| 03 | M | 11 | 15 | K | 11 |
| 08 | H | 11 | 17 | R | 16 |
| 02 | L | 13 | 08 | N | 10 |
| 01 | F | 14 | 11 | Q | 09 |
| 07 | Y | 15 | 18 | W | 15 |
| Odd | 37 | | 43 → | 81-1 | |
| Even | 44 | | 48→ | 91+1 | |
| | 81 | | 91 | | |
| | ______ | | 86-1 | | |
| | - - - - - | | 86+1 | | |
| Odd | 20 | | 52 (68) | | |
| Even | 16 | | 48 (68) | | |
| 06 | A | 04 | 15 | K | 11 |
| 04 | S | 05 | 10 | E | 13 |
| 05 | T | 08 | 07 | D | 12 |
| 03 | M | 11 | 05 | C | 14 |
| 08 | H | 11 | 11 | Q | 09 |
| 02 | L | 13 | 18 | W | 15 |
| 01 | F | 14 | 17 | R | 16 |
| 07 | Y | 15 | 08 | N | 10 |
| Odd | 37 | | 50→ | 86+1 | |
| Even | 44 | | 41→ | 86-1 | |
| | 81 | | 91 | | |
| | ______ | | 76+2 | | |
| | - - - - - | | 96-2 | | |
| Odd | 20 | | 48 (68) | | |
| Even | 16 | | 52 (68) | | |

**Table 3 [left].** The Table follows from Table 1, so that the left octet is given here in a sequence of molecule sizes. In spite of the fact that the sequence of ordinal numbers is disrupted, the odd/even quantums are the same as in Table 1, that means: 16, 20, 48 and 52, but in a vice versa arrangement.

**Table 4 [right].** This Table follows from Table 2 at the same manner as the Table 3 follows from Table 1. Notice that 76 equals 86 – 10 and 96 equals 86 +10.

The amino acid pairs in odd row positions, in Table 1, are larger than the pairs in even row positions. (For example, the amino acid pair F-Q with 25 atoms is larger than L-N with 21 atoms; then the amino acid pair M-K with 26 atoms is larger than S-D with 12 atoms, etc.) Notice also that atom number within side chains of “odd” AAs is determined with multiples of number 6 (8 x 6 = 48 and 9 x 6 = 54). [Cf. these two quantums (48 and 54) with the same quantums in Figure 2.1 in source paper.]

| | | | | | |
|---|---|---|---|---|---|
| 01 | F | 14 | 15 | Y | 07 |
| 02 | L | 13 | 04 | A | 06 |
| 03 | M | 11 | 05 | C | 14 |
| 04 | S | 05 | 08 | T | 05 |
| 08 | H | 11 | 18 | W | 15 |
| 09 | Q | 11 | 08 | N | 10 |
| 11 | K | 15 | 17 | R | 16 |
| 12 | D | 07 | 10 | E | 13 |
| Odd | | 50+1 | 56-1 | | |
| Even | | 36±0 | 30±0 | | |
| | | 86+1 | 86-1 | | |
| | | ——— | 81 | | |
| | | - - - - - | 91 | | |
| Odd | | 23 | 52(68+11) | | |
| Even | | 27 | 34(68-11) | | |

| | | | | |
|---|---|---|---|---|
| F | 14 | | 15 | Y |
| L | 13 | | 04 | A |
| M | 11 | | 05 | C |
| S | 05 | | 08 | T |
| H | 11 | | 18 | W |
| Q | 11 | | 08 | N |
| K | 15 | | 17 | R |
| D | 07 | | 10 | E |
| 50-1 | | Odd (96-2) | 46-1 | |
| 36+2 | | Even (76+2) | 40 | |
| 86+1 | | | 86-1 | |
| | | ——— 91-2 | | |
| | | - - - - - - 81+2 | | |

**Table 5 [left].** The choice of non-contact AAs from GCT. Each next amino acid chooses its own pair-member from all four columns of GCT; on the other words, each amino acid from the left octet is chosen together with its chemically corresponding pair-member in the right octet. Except a determination with quantums “81” and “91” there is a determination with “half” quantums (81 – 1):2 = **40** and (91 + 1):2 = **46** in forms: 40±10 and 46±10. In spite of the fact that the sequence of ordinal number is disrupted, the odd/even quantums (sums) are only in a balance change (±11)
**Table 6 [right].** All is the same (*mutatis mutandis*) as in previous Table, except the pairing process which is arranged horizontally as well vertically. This Table is analog with source paper-Tables 1.1 and 2.1.

| | | | | | |
|---|---|---|---|---|---|
| L13 | A04 | Y15 | F14 | | 46 (36+10) |
| S05 | T08 | C05 | M11 | | 29 |
| Q11 | N08 | W18 | H11 | | 48 (58-10) |
| D14 | E15 | R17 | K18 | | 49 |
| | | | | | |
| 36 | 30 | 56-1 | 50+1 | | 86+1 / 86-1 |
| L13 | A04 | Y15 | F14 | G01 | 47 |
| S05 | T08 | C05 | M11 | I13 | 42 |
| Q11 | N08 | W18 | H11 | P08 | 56 |
| D14 | E15 | R17 | K18 | V10 | 59 |
| V10 | I13 | G01 | P08 | | 102+1 / 102-1 |
| 46 | 46-3 | 56 | 56+3 | | 102 / 102 |

**Table 7 [left].** This Table follows from Table 6 at the same manner as source paper-Figure 1 follows from source paper-Table 1.1. (Cf. quantums 29, 36, 49 and 58 in source paper-Fig. 1.)
**Table 8 [right].** The same is valid as for previous Table; except, the contact AAs are added.

| | | | |
|---|---|---|---|
| 04S | 05 | 08 | T05 |
| 12D | 07 | 10 | E13 |
| 03M | 11 | 05 | C14 |
| 09Q | 11 | 08 | N10 |
| 08H | 11 | 18 | W15 |
| 02L | 13 | 04 | A06 |
| 01F | 14 | 15 | Y07 |
| 11K | 15 | 17 | R16 |
| Odd | 40+1 | 46±0 | |
| Even | 46±0 | 40-1 | |
| | 86+1 | 86-1 | |
| ______ | 81-1 | | |
| - - - - - | 91+1 | | |
| Odd | 16 - - | 41(68-11) | |
| Even | 34 - - | 45(68+11) | |

| | | |
|---|---|---|
| S 05 | | 08 T |
| D 07 | | 10 E |
| M 11 | | 05 C |
| Q 11 | | 08 N |
| H 11 | | 18 W |
| L 13 | | 04 A |
| F 14 | | 15 Y |
| K 15 | | 17 R |
| 36<br>50+1 | Odd (86-10)<br>Even (86+10) | 40<br>46-1 |
| 86+1 | | 86-1 |
| ______ 81 | | |
| - - - - - - 91 | | |

**Table 9 [left].** This Table follows from Table 5 (as Table 3 from Table1).
**Table 10 [right].** This Table follows from Table 9 at the same manner as Table 6 from Table 5.

The sum of two (above mentioned) quantums in Table 1 equals: 48+54 = 102 atoms within 8 molecules, where 102 represents a half of total atom number within 20 canonical AAs, i.e. within their side chains. On the other hand "even" AAs in Table 1 possess 70 atoms which quantum together with the quantum of 32 atoms, existing in four contact AAs (G 01+ P 08 + V 10+ I 13 = 32) equals 102 atoms still once; 102 atoms within 12 molecules. (Proportion 1:1 for atom number and 2:3 for molecule number.) [For proportion 1:1 cf. Marcus (1989) and Stakhov (1989); for proportion 2:3 cf. Moore (1994).] All other illustrations follow analogously and logically – next from previous – as it is shown in their legends.

| | | | | | |
|---|---|---|---|---|---|
| D07 | E10 | T08 | S05 | 30 | |
| Q11 | N08 | C05 | M11 | 35 | 91 |
| L13 | A04 | W18 | H11 | 46 | 81 |
| K15 | R17 | Y15 | F14 | 61 | |
| | | | | | |
| 46 | 40-1 | 46 | 40+1 | 86-1 / 86+1 | |
| D07 | E10 | T08 | S05 | V10 | 40 |
| Q11 | N08 | C05 | M11 | I13 | 48 |
| L13 | A04 | W18 | H11 | P08 | 54 |
| K15 | R17 | Y15 | F14 | G01 | 62 |
| V10 | I13 | G01 | P08 | | 102/102 |
| 56 | 52 | 47 | 49 | | 102+1/102-1 |

**Table 11 [left].** This Table follows from Table 10 at the same manner as source paper-Figure 1 follows from source paper-Table 1.1.
**Table 12 [right].** The same is valid as for previous Table; except, the contact AAs are added.

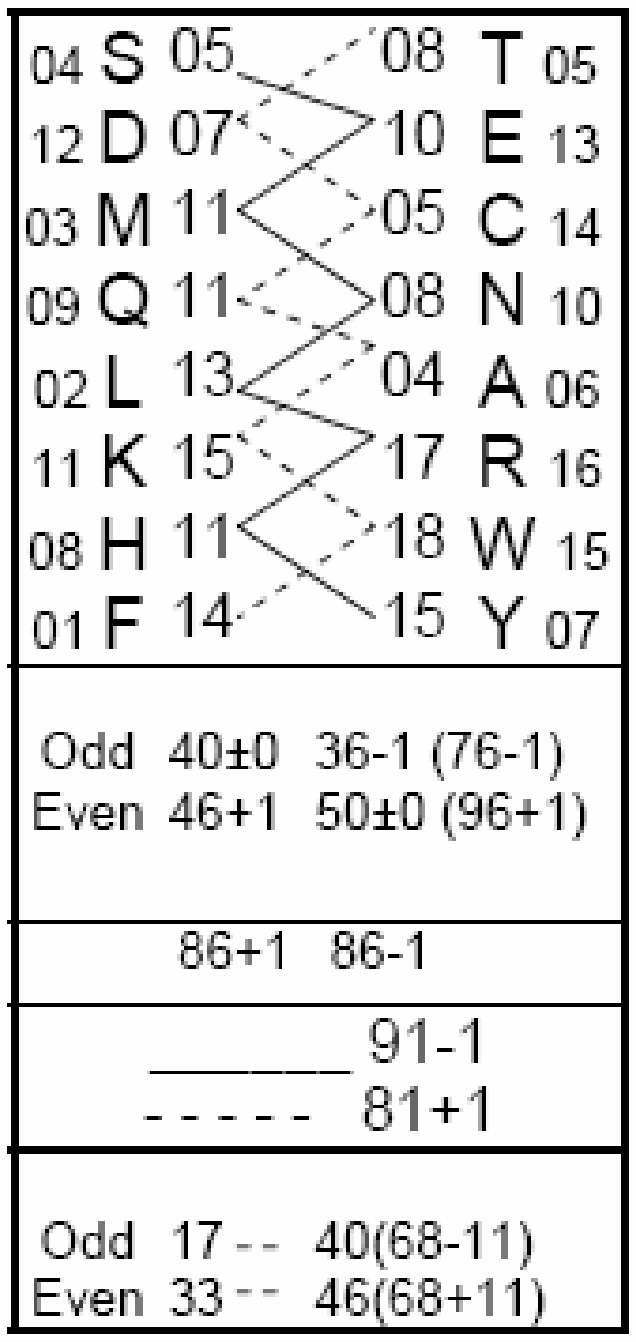

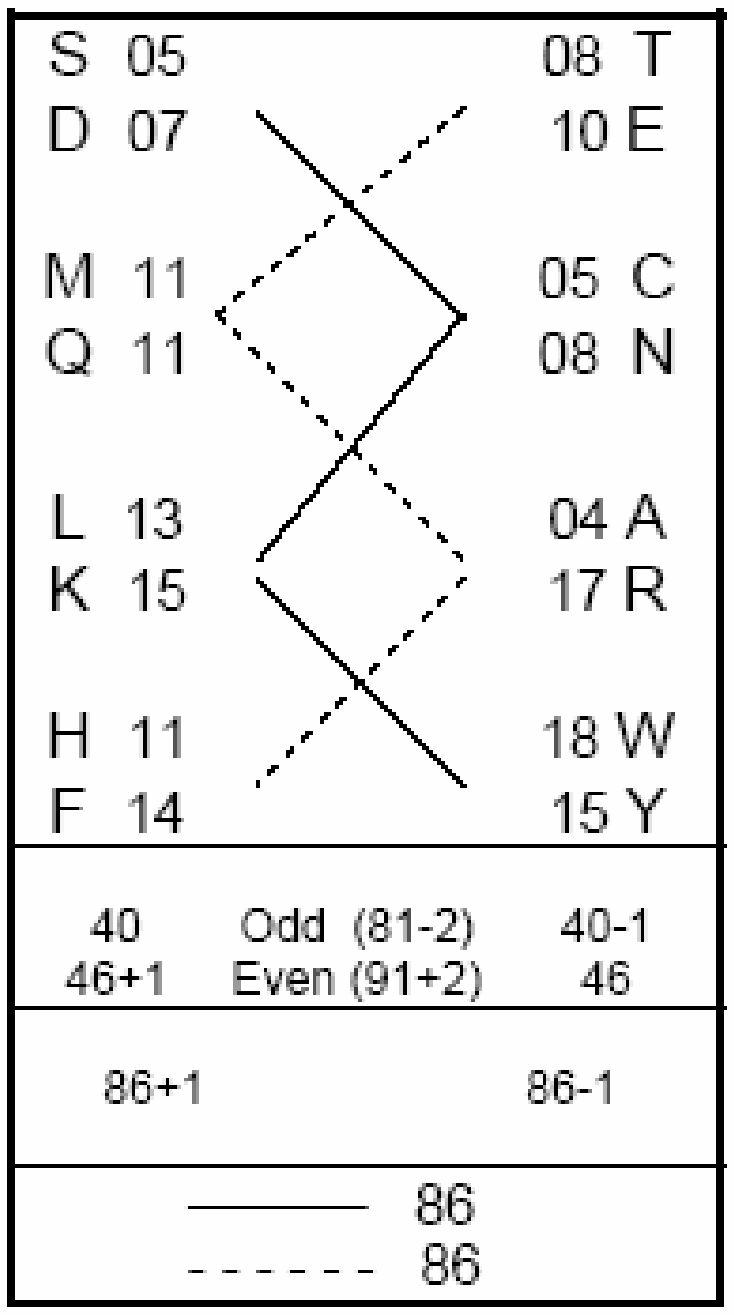

**Table 13 [left].** The order as in Table 5 except a distinction into aliphatic versus aromatic AAs (the first must be aliphatic AAs as in source paper-Table 1.1)
**Table 14 [right].** This Table follows from Table 13 at the same manner as Table 10 follows from Table 9.

| | | | | | |
|---|---|---|---|---|---|
| D07 | E10 | T08 | S05 | | 30 — 91 |
| Q11 | N08 | C05 | M11 | | 35 |
| K15 | R17 | A04 | L13 | | 46 — 81 |
| F14 | Y15 | W18 | H11 | | 61 |
| | | | | | |
| 46+1 | 50 | 36-1 | 40 | | 86+1 / 86-1 |
| D07 | E10 | T08 | S05 | **V10** | 40 |
| Q11 | N08 | C05 | M11 | **P08** | 43 |
| K15 | R17 | A04 | L13 | **I13** | 62 |
| F14 | Y15 | W18 | H11 | **G01** | 59 |
| **I13** | **V10** | **P08** | **G01** | | 102 / 102 |
| **60** | **60** | **43** | **41** | | **102+1 / 102-1** |

**Table 15 [left].** This Table follows from Table 14 at the same manner as source paper-Figure 1 follows from source paper-Table 1.1.
**Table 16 [right].** The same is valid as for previous Table; except, the contact AAs are added.

## SUPPLEMENT 2

**1. Introduction.** In this supplement we give still a set of harmonic structures of the genetic code, determined by Gauss arithmetical algorithm as it is shown in the source paper (Rakočević, 2006a), q-bio.OT/0610044. In source paper beside others, we showed that polarity of amino acids (AAs) indirectly is determined by Gauss arithmetical algorithm. Namely, the polar AAs are positioned (within a 4 x 5 amino acid system) as a separate entity, in the form of a specific "island" surrounded by non-polar AAs (source paper – Figure 2.1). By this, the four ambivalent AAs, i.e. polar and nonpolar at the same time (glycine, proline, tryptophan and histidine)[1] are positioned in a snug "string" at the very edge of the system.

In this supplement, however, we show a direct determination, i.e. a direct connection between Gauss arithmetical algorithm and polarity of amino acid molecules, including their positions in standard Genetic Code Table (GCT). In such a case the connection itself appears to be followed by a strict atom number balance through the existence of new four amino acid classes.

**2. Results.** The starting step is the choice of one and only single amino acid pair, G-P (Table 1); single pair because glycine is one and only amino acid within glycinic stereochemical type and proline also one and only, but within prolinic stereochemical type (about four stereochemical types see in Popov, 1989 and Rakočević & Jokić, 1996). Follow two pairs of aromatic AAs, first H-W and second F-Y. As the first pair is H-W because the quartet G-P/H-W represents ambivalent AAs – polar and nonpolar at the same time (cf. Footnote 1). The next steps are comming in relation to the AAs sequence, such as it occurs in the first column of GCT.

As a noteworthy is the fact that after four ambivalent come four extrem AAs, extrem just from the aspect of polarity: F-Y/L-R. Phenylalanine is the most nonpolar amino acid of all four aromatic AAs and Leucine (together with isoleucine) is the most nonpolar of all aliphatic AAs (after hydropathy). On the other hand, Tyrosine is extrem because its polaryty comes not only from the aromatic ring but from a polar functional group (OH); and arginine is extrem through its very massive guanidino group – very massive, and very polar at the same time.

---

[1] Glycine: after hydropathy is polar; after cloister energy and polar requirement is non-polar; Proline: after hydropathy and cloister energy is polar; after polar requirement is non-polar; Tryptophan: after hydropathy and polar requirement is polar; after cloister energy is non-polar. Hydropathy (Kyte & Doolittle, 1982); cloister energy (Swanson, 1984); polar requirement (Woese et al., 1966; Konopel'chenko and Rumer, 1975). Really regarding, histidine as a „semi-ambivalent" amino acid since it has neither positive nor negative value in cloister energy, but its value is equal to zero (Figure 5 in Swanson, 1984). About the pairing process of AAs through Hydropathy and Cloister energy *see* Survey 1 in Rakočević & Jokić, 1996, and about ambivalence of glycine and proline see Section 3.3 in Rakočević, 2004b.

As we can see from equations 1-4 and from Table 2, the four extreme AAs are the same AAs that we know from the balance distinction between **p**olar/**n**onpolar[2] and **i**nner/**o**uter AAs in GCT, with the reading: 22 molecules, 222 atoms within amino acid side chains and 420 atom within whole molecules, etc. (Rakočević, 2000, 2006c):

(n) 4V+1M+3I+4A+**2L**+**4L**+**2F**+2C = 22
40+ 11+39+16+26 + 52 +28+10 = 2**2**2 (42**0**) (1)

(o) 4V+1M+3I+4A+**2Y**+**4R**+**1W**+2C = 21
40+ 11+39+16+ 30 + 68 +18+ 10 = 2**3**2 (42**1**) (2)

(p) 4G+2K+2N+4P+**2Y**+**4R**+**1W**+2E+2D+4T+2R+2S+2Q+2H+4S = 39
04+30+16+32+30 + 68+ 18+20+14+32+34+10+22+22+20 = 3**7**2 (72**3**) (3)

(i) 4G+2K+2N+4P+**2L**+**4L**+**2F**+2E+2D+4T+2R+2S+2Q+2H+4S = 40
04+30+16+32+26+52 + 28+20+ 14+32+ 34 +10+22+ 22+ 20 = 3**6**2 (72**2**) (4)

(p) 6S+4T+2N+2Q+2D+2E+2K+6R+2Y = 28
30+32+16+ 22+ 14+20+ 30+102+30 = 297–1 (549–1) (5)

(n) **4G**+**4P**+**1W**+**2H**+4A+6L+4V+3I+2C+1M+2F = 33
04+32+ 18+ 22+ 16 + 78 +40 +39+10+11+ 28 = 297+1 (594+1) (6)

Equations (5) and (6) show the relations which are more than a balance; a specific relationship of the amino acid "heads" and "bodies" (all "heads" and all "bodies") from one side and the wholeness of molecules from the second side. So, the distribution of atom number is the following. In polar AAs molecules there are exactly as many atoms as in the heads of all molecules, minus one atom (61 x 9 = 549). In the "nonpolar" molecules (nonpolar plus ambivalent) there are as many atoms as in the bodies, i.e. side chains of all 61 molecules, plus one atom [(297+1) + (33 x 9 = 297) = 594 +1]. On the other hand, in the bodies of polar as well as of non-polar molecules, there is 297±1 atoms, exactly as in first (297-1) and second half (297+1) of standard GCT[3]. Thirdly, in the heads of non-polar molecules there are exactly as many atoms as it is the half of the atom number in the bodies of all molecules (33 x 9 = ½ 594), while in the heads of polar AAs there are as many as it is the half of 5**0**4, which number is the modular pair (in module 9) of number 5**9**4.

[2] Polar/nonpolar AAs after their hydropathy (cf. Footnote 1).

*Remark* 1. Number 504 represents the sum of each two and two (out of eight in total) branches on the 6-bit binary tree (Figure 1 in Rakočević, 1998), or of two and two octets in GCT as follows: in two central octets there are: 24+25+ ...+31 = **220** and 32+33+ ...+39 = **284** (220+284 = 504); in next two octets there are: 16+17+ ...+23 = 156 and 40+41+ ...+47 = 348 (156+348 = 504); within the first to last pair of octets there are: 8+9+ ...+15 = 92 and 48+49+ ...+55 = 412 (92+412 = 504); Finely, within the last octet pair we have: 0+1+ ...+7 = 28 and 56+57+ ...+63 = 476 (28+476 = 504)(cf. Table 3).

*Remark* 2. The number 504 represents the sum of the first two friendly numbers (the first pair): 220+284 = 504. On the other hand, Shcherbak has shown (1994) that within the set of 23 AAs, the eight four-codon AAs possess 592+333 = 925 nucleons, where 592 is a half of the third friendly number (1184 = 2 x 592)[4]. But Shcherbak also showed that within side chains of all 23 AAs there are 1443 of nucleons (333 nucleons within four-codon AAs plus 1110 within non-four-codon AAs), where 1443 is a sixth part of the sum of first four perfect numbers [1443 x 6 = 8658 = (7770+0888) = 6+28+496+8128] (About determination of the genetic code with the perfect and friendly numbers see in Rakočević, 1997).

Parallel with the atom number balance there is a molecule number balance as follows: the unit distances between the number of molecules: 22-21 = 1 and 40-39 = 1 in equation (1) in relation to equation (2) and equation (4) in relation to equation (3), respectively; then the double units distances between the number of molecules: 33-22 = 11 and 39-28 = 11 in equation (6) in relation to equation (1) and equation (3) in relation to equation (5), respectively.

---

[3] Table 2 appears to be the standard GCT if the texture (dark tones) is excluded. In such a manner the left half of GCT make 32 amino acid molecules encoded by 32 NYN codons, whereas the right half make 29 amino acid molecules encoded by 29 NRN codons, plus three stop codons. Within 32 amino acid molecules (side chains) there are 297-1, and within 29 molecules 297+1 of atoms.

[4] The forth friendly number is the number 1210 as a product of 10 x $11^2$. The third and forth friendly numbers make the second friendly number pair.

| | | | | | | |
|---|---|---|---|---|---|---|
| G 01 | H 11 | **F** 14 | **L** 13 | **I** 13 | **M** 11 | **V** 10 |
| P 08 | W 18 | Y 15 | R 17 | K 15 | C 05 | A 04 |
| | | | Q 11 | N 08 | S 05 | T 08 |
| | | | E 10 | D 01 | | |

**Table 1**. First four AAs are ambivalent, next four "extreme" (as it is explained in the text) and other AAs in a chemically determined order and arrangement.

After first two AAs quartets in Table 1 (G-P/H-W and F-Y/L-R) follow other AAs through a chemically relevant relation: two source aliphatic AAs (L-I)[5] and two amino derivatives (R-K); then two sulfur AAs (M-C) and two source aliphatic still once (V-A). The next two are the chalcogene AAs: S-T in a continuation to M-C; amino acid T also in contact with A as two methyl derivatives. At the end come two carboxylic AAs (D-E) whose amide derivatives, as nitrogen compounds, hold a connection with other two nitrogen derivatives (R-K).

[5] Notice that F-L make also a chemical pair through the same structural motive – the first possible branching (iso-butane in relation to toluen structural motive). Here lies the reason why benzene ring is axcluded from the set of aromatic AAs.

| 1st lett. | 2nd letter: U | C | A | G | 3rd lett. |
|---|---|---|---|---|---|
| *U* | UUU<br>UUC F II<br>UUA<br>UUG L I | UCU<br>UCC<br>UCA S II<br>UCG | UAU<br>UAC Y I<br>UAA<br>UAG CT | UGU<br>UGC C I<br>UGA CT<br>UGG W I | *U*<br>*C*<br>*A*<br>*G* |
| *C* | CUU<br>CUC<br>CUA L I<br>CUG | CCU<br>CCC<br>CCA P II<br>CCG | CAU<br>CAC H II<br>CAA<br>CAG Q I | CGU<br>CGC<br>CGA R I<br>CGG | *U*<br>*C*<br>*A*<br>*G* |
| *A* | AUU<br>AUC Ile I<br>AUA<br>AUG M I | ACU<br>ACC<br>ACA T II<br>ACG | AAU<br>AAC N II<br>AAA<br>AAG K II | AGU<br>AGC S II<br>AGA<br>AGG R I | *U*<br>*C*<br>*A*<br>*G* |
| *G* | GUU<br>GUC<br>GUA V I<br>GUG | GCU<br>GCC<br>GCA A II<br>GCG | GAU<br>GAC D II<br>GAA<br>GAG E I | GGU<br>GGC<br>GGA G II<br>GGG | *U*<br>*C*<br>*A*<br>*G* |

**Table 2**. The amino acids within three diagonals are inner, and other – outer AAs. On the other side, within bordered space are polar AAs and other – nonpolar AAs (polar/nonpolar after hydropathy: Kyte & Doolittle, 1982).

/00 - 07/08 - 15/16 - 23/24 - 31//32 - 39/40 - 47/48 - 55/56 - 63/
28 92 156 220 284 348 412 476
64 64 64 64 64 64 64

/00 - 07/00 - 15/00 - 23/00 - 31//00 - 39/00 - 47/00 - 55/00 - 63/
28 120 276 496 780 1128 1540 2016
92 156 220 284 348 412 476

**Table 3**. The eight octets within 6-bit binary-code tree (Rakočević, 1998) as well as within GCT are determined with the first pair of friendly numbers (220 & 284) and third perfect number (496). For details see the text, especially Remarks 1 & 2.

| | | | |
|---|---|---|---|
| **G** | 01 | 08 | **P** |
| **H** | 11 | 18 | **W** |
| **F** | 14 | 15 | **Y** |
| **L** | 13 | 17 | **R** |
| **I** | 13 | 15 | **K** |
| **M** | 11 | 05 | **C** |
| **V** | 10 | 04 | **A** |
| **T** | 08 | 05 | **S** |
| **Q** | 11 | 08 | **N** |
| **E** | 10 | 07 | **D** |
| Odd | **49** | **50** | |
| Even | **53** | **52** | |
| | 102 | 102 | |
| **G** | 001 | 041 | **P** |
| **H** | 081 | 130 | **W** |
| **F** | 091 | 107 | **Y** |
| **L** | 057 | 100 | **R** |
| **I** | 057 | 072 | **K** |
| **M** | 075 | 047 | **C** |
| **V** | 043 | 015 | **A** |
| **T** | 045 | 031 | **S** |
| **Q** | 072 | 058 | **N** |
| **E** | 073 | 059 | **D** |
| O | **264** | **293** | |
| E | **331** | **367** | |
| | 595 | 660 | |

**Table 4 [left].** This Table follows from Table 1. Atom number determination in relation to Gauss' algorithm (explanation in the text)
**Table 5 [right].** All is the same as in previous Table, except the determination by nucleon number. Notice a symmetry determination through module 9: 5**9**5 versus 66**0**.

From Table 1 follows Table 4 (in relation to Table 5), first row on the left and second row on the right, plus AAs from third and fourth rows – three outer (T, Q, E) on the left and three inner (S, N, D) on the right. As we see AAs in odd and even positions make four AAs groups with atom number directly determined by Gauss' arithmetical algorithm (Table 6). Namely, in source paper (source paper – Figure 1) we showed that within four amino acid rows there are so many atoms as in 10$^{th}$ and 20$^{th}$ Gauss' pair (in relation to middle point "51": 10$^{th}$ pair as 41-61 and 20$^{th}$ pair as 31-71 atoms). And here we see that within four amino acid rows are so many atoms as in 01$^{st}$ and 02$^{nd}$ Gauss' pair (01$^{st}$ pair as 50-52 and 02$^{nd}$ pair as 49-53 atoms).

| | | | |
|---|---|---|---|
| **G** | P | W | **H** |
| **F** | Y | R | **L** |
| **I** | K | C | **M** |
| **V** | A | S | **T** |
| **Q** | N | D | **E** |
| **49** | 50 | 52 | **53** |
| **G** | M | P | **C** |
| **H** | V | W | **A** |
| **F** | T | Y | **S** |
| **L** | Q | R | **N** |
| **I** | E | K | **D** |
| **52** | 50 | 73 | **29** |

**Table 6 [left].** The Table follows from Table 4: two outer columns as odd/even positions in left column of table 4, and two inner columns as odd/even positions in right column of table 4, respectively.

**Table 7 [right].** As previous one, this Table also follows from Table 4: first five and last five AAs as two outer columns, and 5 & 5 amino acids as two inner columns. The atom number quantums "50" and "52" are the same as in previous Table (Table 6), whereas two other quantums (73 = 71+ 2 and 29 = 31 – 2) correspond to the Gauss pair 31-71 through a deviation of ±2 (minus first and second step; plus first and second step).

The sum of two quantums in Table 1 equals: 48+54 = 102 atoms within 8 molecules, where 102 represents a half of total atom number within 20 canonical AAs, i.e. within their side chains. On the other hand "even" AAs in Table 1 possess 70 atoms which quantum together with the quantum of 32 atoms, existing in four contact AAs (G 01+ P 08 + V 10+ I 13 = 32) equals 102 atoms still once; 102 atoms within 12 molecules. (Proportion 1:1 for atom number and 2:3 for molecule number.) [For proportion 1:1 cf. Marcus (1989) and Stakhov (1989); for proportion 2:3 cf. Moore (1994).] All other illustrations follow analogously and logically – next from previous – as it is shown in their legends.

**3. Conclusion for both Supplements.** Bearing in mind that the order of amino acids in presented Tables, in both Supplementsis, is given in correspondence with the order of codons in GCT, it makes sense to speak about genetic code as a harmonic system. On the other side, presented harmonic structures provide evidence to support the hypothesis, given in a previous paper (Rakočević, 2004b), that genetic code was complete from the very beginning as the condition for the origin and evolution of the life.

## SUPPLEMENT 3. *The Cipher of the Genetic Code*

**1. Introduction.** In this Supplement is presented a new approach to understanding of the genetic code. In order to overcome the key paradox (and Darwinian selection problem) that the highly complex amino acid Phe is encoded by the simplest codons (UUY), and the simplest Gly encoded by the most complex codons (GGN); as well as the paradox of the duplication of some amino acids in the encoding process (Leu, Ser, Arg), we proposed an extension of the notion (and concept) of genetic code. For a better (and lighter) understanding of genetic coding, we proposed a hypothesis after that (under the conditions of allowed metaphoricity and modeling in biology) genetic code has to be understood, analogously to understanding in cryptology, as the unity of the three entities: the code, the cipher of the code and the key of the cipher. In this hierarchy the term (and notion) "genetic code" remains what has been from the beginning: a connection between four-letter alphabet (four Py-Pu nucleotides, in form of codons) and a twenty-letter alphabet (twenty amino acids); the cipher is a specific chemical complementarity in chemical properties of molecules in the form: similarity in dissimilarity versus dissimilarity in similarity ("Sim in Diss vs Diss in Sim") and the key of cipher: the complementarity on the binary tree of the genetic code in the form: 0-15, 1-14, 2-13, ..., 6-9, 7-8. Just only with this understanding, it appears a possibility for an additional understanding that within the two main Genetic Code Tables (of the nucleotide doublets and Triplets) exists a sophisticated nuancing and balancing in the properties of th constituents of GC, including the balance of the number of molecules, atoms, and nucleons.

[**Nota bene**: "Before discussing these problems ..., we must address a preliminary one. We must face the *ontological problem* of the reality of the organic codes: are they real codes? Do they actually exist in living systems? It is a fact that the genetic code has been universally accepted into Modern Biology, but let us not be naive about this: what has been accepted is the *name* of the genetic code, not its *ontological reality*. More precisely, the genetic code has been accepted under the assumption that its rules were determined by chemistry and do not have the *arbitrariness* that is essential in any real code. The theoretical premise of this assumption is the belief that there cannot be arbitrary rules in Nature, and this inevitably implies that the genetic code is a metaphorical entity, not a real code. This idea has a long history and let us not forget that for many decades it has been the dominant view in molecular biology" (Marcello Barbieri, 2018, p. 2).][6]

**2. A possible scenario for molecules selection.** With the revelation of the GCT, in both forms as the Table of the nucleotide triplets (Crick, 1966) and the Table of nucleotide doublets (Rumer, 1966), it has become an obvious scenario by which the chemical constituents of the genetic code are generated; both builders, of nucleic acids, as well as of protein amino acids. The first thing that is directly visible is the fact that all constituents (molecules) of genetic code are built from the first few simplest elements (non-metals) at the beginning of the Periodic System of Chemical Elements (PSE). From the first four elements (H, C, N, O), pyrimidine (Py) and purine (Pu), that is their genetic code derivatives, were constructed; also the 18 AAs. [These four elements are in immediate neighborhood, three in continuity (C, N, O), at the beginning of the 4$^{th}$**,** 5$^{th}$ and 6$^{th}$ groups, respectively, and H at the beginning of the 7$^{th}$ group, as the diagonal neighbor of oxygen.] To complete the nucleotide molecules, and to build two more AAs, which have sulfur, the Darwin's selection sieve must expand its openings in order to select phosphorus (as the vertical neighborhood of nitrogen) and sulfur (as the vertical neighborhood of oxygen) from the next period of PSE.

By the "chemical eyes" seeing, three start precursor molecules are: benzene, methane and cyclopropane, the simplest hydrocarbons from the corresponding groups – arenes (the most stable hydrocarbons), alkanes and cycloalkanes, respectively. The potential candidates for builders are themselves, or their derivatives. Bearing in mind chemical formulas, we find that aromatic hydrocarbons provide precursors for all four Py-Pu bases and all four aromatic AAs – the simplest aromatics, from the most stable six-membered and five-membered groups. On the other hand, observing the structure of the methane molecule, we note that here (in the act of selection), besides the principle of similarity, the principle of self-similarity applies: methane structure pattern is not only the pattern included into the "head" (amino acid functional group) of each of 20 AAs, than also in the body of all 16 AAs of the alanine stereochemical type. [Beside that the methan structural pattern (through the $CH_2$ group, located between the head and the "body", i.e. side chain of AA) makes the basis of the alanine stereochemical type, the glycine type possess also the same pattern.][7]

[6] "The very first **model** of the genetic code was the *Stereochemical Theory*, an idea proposed by George Gamow in 1954 ... The second **canonical model** was the *Coevolution Theory* proposed by Wong (1975, 1981), according to which the genetic code coevolved with the biochemical pathways that introduced new amino acids in protein synthesis" (Barbieri, 2018, p. 2) (The bolding: MMR).

Through the structure of alanine, the simplest amino acid of the alanine stereochemical type, all four aromatic AAs are included in a set of 16 AAs of alanine type. Phenylalanine, as its name suggests, is an alanine derivative by replacing a hydrogen atom in the side chain of alanine, in the $CH_3$ group, with a benzene phenyl group. By this act appears the situation which is also readable as so that phenylalanine is formed as a derivative of derivative: in the benzene derivative toluene one hydrogen atom is replaced by an amino acid functional group. All together, the self-similarity of amino acid molecules in them-selves is realized.[8]

After the selection of phenylalanine (at the beginning of the first column of GCT), as the first possible aromatic AA with a six-membered ring, its first possible derivative, tyrosine, is selected (at the beginning of the third column of GCT). It possesses the hydroxy group in the most stable para position (and not in a less stable ortho or meta position). Following two Mendeleev principles – the continuity and minimum of change – one should expect a derivative with a nitrogen functional group (amino group), instead of the oxygen (hydroxyl) group. But in such a case we would have a functional group with three instead with two atoms, what would not be in accordance with the principles of balancing and nuancing, which, in the case of genetic code, are also valid, as we will further show.

If we see the beginning and the end of GCT as complementary to each other (see below), then we see that, as the first minimal change in the set of arenes occurs at the beginning, so does the very first minimal change in the set of alkanes. [It is, therefore, the complementarity of the two most stable classes of hydrocarbons; one that is very complex, with hybridized chemical bonds and delocalized electron orbitals (arenes) and others that are very simple, with simple chemical bonds and localized electron orbitals (alkanes).][9]

---

[7] About four stereochemical types of AAs one can see in (Popov, 1989; Rakočević & Jokić, 1996).

[8] Rakočević, 2004b, p. 231: . "Hypothesis on a [prebiotically] complete genetic code (CGC): By this hypothesis ... we support the stand point that CGC must be based on several key principles. ... 1. The principle of systemic self-related and self-similar organization. …" [Note: This hypothesis best corresponds with the same such hypothesis of V.V. Sukhodolets (1985).]

As in the beginning, a minimal change occurs in which from Phe follows Tyr, so the same or similar process occurs at the end: the substitution of a hydrogen atom in Gly (last in the fourth column), with one methyl group follows Ala (last in second column), which is "automatically" (as explained above) in relation to Phe. Moreover, the substitution of hydrogen atom in the Gly with an isopropyl group appears Val (last in the first column). At the "same time", together with valine was generated proline; and in a parallel process, in the first row, as a direct derivative of alanine, follow Ser, the first in second column, and Cys in the fourth one.

Already with the selection of the first aromatic AA (Phe), we see the correspondence with two Py nucleotide bases, because the pyrimidine is a benzene derivative. As can be seen from molecules formulas, the selection of a two-nitrogen pyrimidine is preferable than one-nitrogen pyridine. Two chemical reasons can be crucial here. Pyrimidine is more than a weaker base, but, more importantly, its far greater ability is to establish hydrogen bonds in potential dimers, which are actually found in natural DNA and RNA. One and the other is a distinct advantage for further balancing and nuancing. [As the "sowing" through the Darwin sieve, in the selection of chemical elements of the second period of the PSE, stopped before the fluorine, which, with its high reactivity, "burns" life before it arises, so this has also been shown here that the sieve was non-selective for stronger organic bases, which also lack the ability for balancing and nuancing in dimerization.]

With the selection of the third (Trp) and the fourth (His) aromatic AAs, the correspondence with two Pu nucleotide bases is also evident. In addition, it is also evident that in Darwin's prebiotic (chemical) evolution the selectivity of the sieve is "enriched" so that two additional principles apply: the principle of economicity and the principle of the enriching of diversity by increasing the degree of multi-meaning in relationships. The influence of both these principles is seen in the act of selection of tryptophan. In the case of Trp, we have a fusion of benzene with pyrrole into a two-ring indole, rather than with pyridine in double-strand quinoline; in the case of histidine, imidazole is selected as similar to pyrrole.

[9] "Contraria sunt complementa." (A motto at Niels Bohr's own coat of arms, which featured a taijitu, symbol of yin and yang, designed in 1947.

In addition to the stated reasons for the non-ability of pyridines, the reason are also the two quoted principles: by selection a five-membered aromatic ring (in both cases, in the case of pyrrol, as well as imidazole), instead of the six-member, molecular diversity increases, and in addition, imidazole provide an aromatic electronic sextet which possess the benzene too; moreover, the selection of this new type of electronic sectet enriches the multi-meaning relations within the genetic code.

Both five-membered rings, pyrrole and imidazole, in fused compounds, within the constituents of genetic code, provide approximately equal acidity/basicity with a lower reactivity. Imidazole is found in purine, in two purine bases, but also, as already mentioned, in the fourth aromatic AA histidine (the fourth in this discussion).

Additionally, one more point is needed here. When we discuss the analogy of six-membered and five-membered aromatic rings, apart from the similarities presented, we also mean the coherence of structural patterns (structural motives) and the similarity in the "flow" of delocalized electrons and electronic densities in the molecule. Thus, for pyrimidine and imidazole, except for the possession of an aromatic sextet, we say that they are analogs with the fact that they both have non-adjacent nitrogen atoms. [The pyrimidine is not analogous to the pyrazole (the imidazole isomer) because both the pyrazole nitrogen atoms are adjacent.]

Above we analyzed that part of the scenario that relates to the selection of Py-Pu bases and four aromatic AAs. Now we are going to analyze the generating, i.e. the selection of those AAs, which have a "pure" hydrocarbon side chain, a standard series (Ala, Leu, Val, Ile) and two AAs with a non-standard side chain (Gly, Pro). We assume that the logic of selection by the principle of matching the same or similar structural patterns here is also valid. Through analyzing these six AAs, we will continue to "keep on eye", for comparison, the remaining eight non-sulfur AAs.

*

Observing the structure of methane, the simplest alkane, we see that it is a form that we find in the amino acid functional group (in the "head" of AA), as already mentioned above. Even more than that, we note that the generating of the body (side chain) of the remaining eight AAs can be understood as a "mapping" of partial functional groups from the head to the body. And, the mappings are as follows: the amino group leads to the generating of Lys & Arg; hydroxyl to Ser & Thr; carboxyl to Asp & Glu; and, finally, the fusion of the carbonyl and amino groups forms two amides (Asn & Gln), the derivatives of the two carboxylic AAs.

*

We return to the problem of generating four standard hydrocarbon AAs (Ala, Leu, Val, Ile) and two non-standard (Gly, Pro). We continue to observe the structure of the methane molecule and note that this structure does not resemble the structure of the following members of the homologous series of alkanes (ethan and propane). Paradoxically, the methan looks like the fourth member, not as n-butane but as iso-butane. Despite the fact that, after Ala, as methyl derivatives, it is expected that AAs, which are ethyl and propyl derivative, are selected; this, however, does not happen; such AAs exist not in the set of protein AAs.

The only difference between methyl and iso-butyl group is the size of electronic density on three "branches". Therefore, it becomes understandable why, after alanine, with the simplest hydrocarbon sequence of one methyl group, the expected derivatives with the atomic groups of ethyl, n-propyl, n-butyl come not, but iso-propyl and isobutyl derivatives, as well as one nitrogen analogue – guanidine molecule. [The analogy through the iso-propyl group and the guanidine molecule.]

Isobutane derivatives are the only two isomeric AAs, leucine and isoleucine; Leucine as the second AA in the series of AAs of the alanine stereochemical type and isoleucine as a second one in series of AAs of the valin stereochemical type. [The generating of the first AA in the alanine type (Ala) on the methane pattern was explained above; also of the generating of the first, and only one AA in the glycine type (Gly). The explanation, however, for the first AA in the valine type (Val) and the first, and only one in proline type (Pro) on the isopropil pattern, we gave in one of the previous papers (Rakočević & Jokić, 1996).]

**3. A specific chemical complementarity as the cipher.** Bearing in mind the Gray Code model of the genetic code (Swanson, 1984), Binary-code tree (Rakočević, 1998) and the Scenario, presented in previous Section, just their possible connection, a new arrangement of genetic code can be generating (Table 1). Such an arrangement makes sense to be named CIS (Canonical Invariant System) since the canonical amino acids in it are strictly determined and possess positions which it cannot be changed.

In the next step it makes sense to bring together the CIS and the Rumer's Table of nucleotide doublets (Rumer, 1966), where the one-meaning doublets will be separated from the two-meaning ones (Table 2); such a procedure in order to analyzing the interrelations of splitted doublets from the aspect of two types of arrangements: according to their positions in the CIS and according the number of their hydrogen bonds.

The specificity of the chemical complementarity, which we are talking about here, is expressed as complementarity of the nucleotide doublets as well as the corresponding amino acids on each two vertices on the Boolean hypercube ($B^4$) whose binary values give the sum 1111 in the binary record (corresponding to the number 15 in decimal one). In the question are the nuanced and balanced chemical complementarities, both through chemical structures, and through the chemical properties of molecules. Examples are already listed in the Scenario, and here we again present some characteristic examples in the new meaning (cipher – key of the cipher). Thus, we see that the initial (zeroth) doublet UU is complementary to the last doublet GG, which is the complementarity through the *chemical dissimilarity*.

*

Of the two pyrimidines, uracil is simpler because it has two the same functional groups (oxo functional groups), while cytosine possesses two different functional groups – oxo and amino. Of two purines, glutamine is more complex because it possesses both of these groups, while adenine has only one amino group. Paradoxically, the simplest doublet, UU, encodes Phe, very complex AA, in whose side chain is very complex benzene ring. On the other hand, the most complex doublet, GG, encodes Gly, the simplest within the set of 20 protein AAs, with only one hydrogen atom in side chain. Thus, we might say that this complementarity is the complementarity through dissimilarity. However, a deeper chemical insight shows that this complementarity is a complementarity through *similarity in dissimilarity*. In the case of two dissimilar nucleotides (as Py vs Pu), UU vs GG, the similarity is that both possess the oxo group; in the case of two dissimilar AAs (Phe vs Gly), the similarity is in a similar structural pattern within both molecules. The matter becomes clearer when it is seen that the doublet UU does not encode only Phe than Leu. [In reality the codons code for amino acids. Thus, Gly is encoded with the four most complex codons (GGN). On the other hand, chemically very complex AA, Phe, is encoded with only two simplest codons, UUU & UUC.]

Within the $CH_3$ group of toluene, one hydrogen atom is substituted by an amino acid functional group. Hence, a H-C-H group is formed between the head and the body. In this case, a C atom from the amino acid functional group and a C atom from the benzene ring are bonded vertically for the C atom in H-C-H group. All together, a form analogous to structures of methyl and isobutyl group is obtained.

As we see, in the case of nucleotides, the dissimilarity is derived from the difference between their two dissimilar types (pyrimidine vs purine), and the similarity is derived from the same functional group. In the case of AAs, the dissimilarity arises from the difference of molecular classes involved in the construction of the amino acid side chain, and the similarity commes from the similarity in molecular structural patterns.

**4. A neighborhood logic.** In order to be able to analyze other examples of complementarity in the standard GCT, we must first answer the question of which GCT arrangement must be presented; in other words, which of 72 possible the arrangements, is the best; the best in terms of respecting the chemical hierarchy, from the aspect of less or more molecular complexity. The correct answer to the raised question was already given in 1966 in the form of the Nucleotide Triplet Table (Crick, 1966, 1968) and the Nucleotide Doublets Table (Rumer, 1966). And what was in force in 1966 and 1968, is valid also today, with all in the presence the so-called deviant codes (Box 1).

The nuancing and balancing of chemical structures (and properties) through complementarity becomes obvious only if we in the first and second columns of the GCT go downward, and in the last and the first to last – upward. By this act, there is an immediate obviousness that the "neighborhood logic" of codons and correspondent AAs, valid for columns, extends also to rows. Even more than the neighborhood logic also applies in diagonal connections. The examples are as follows. Phe and Leu are (through complementarity logic) in connection with Gly, as we said above. However, according to the logic of the neighborhood, they are at the same time in connection with Ser and Pro; Leu even more with Thr. In Phe the bonding between the head and the body was achieved on a methane and isobutane pattern; In Leu, the body itself is that pattern. In the Ser we have the first possible derivation of the methan pattern by the introduction of the hydroxyl group. This introduction of a hydroxyl group chemically can also be understood as the substitution of one hydrogen atom in the $CH_3$ group of Ala.

**Box 1.** *The deviant genetic codes*
At that very first time (from 1966 until 1979) the Genetic Code Table was considered to be the Table of a universal genetic code. However, later with the discovery of alternative genetic codes, the Table was renamed in GCT of standard genetic code. The universality of the genetic code was first challenged in 1979, when mammalian mitochondria were found to use a code that deviated somewhat from the "universal" (Barrell et al., 1979; Attardi, 1985). Our opinion about "deviant codes" we have expressed in one of the previous works (Rakočević, 2004), and we still think the same today, that they represent only a "degree of freedom" in deviation from the standard one. [Knight at al., 2001, p. 49: "The genetic code evolved in two distinct phases. First, the 'canonical' code emerged before the last universal ancestor; subsequently, this code diverged in numerous nuclear and organelle lineages"; Weaver, 2012, pp. 568-569: "These deviant codes are still closely related to the standard one from which they probably evolved".]

A "step further" is realized in the Thr – it is formed by replacing one hydrogen atom in the $CH_2$ group of the Ser by a methyl group. Hence, in the Thr between the head and the body is not H-C-H, but H-C-$CH_3$ group. This also explains the similarity of Thr, both with Ala and with Pro, what corresponds to their positions in the GCT. In the Pro, the side chain consists of three methylene ($CH_2$) groups, analogous, in its commonality, as a sequence of three such groups, with the structure of the isopropyl group. ["... just as a $–CH_2–$ group in a molecule is called a methylene group" (Wade, 2013, p. 166).] By respecting the principle of self-similarity, the side chain of Pro can be chemically "read" as a pyrrolidine, which is a non-aromatic analogue of pyrrole.

After phenylanaline, going from the left to the right in the rows, we find a very characteristic case of simultaneous linear and diagonal neighborhoods between the three remaining aromatic AAs: Tyr, His and Trp. After phenylanaline and leucine in the first column, isoleucine and methionine come in a complementarity relationship with arginine. In other words, when we go upstairs, in the last column, we encounter the arginine for which we saw that in the side chain it possesses a structural motif analogous to the structural motif of the isolpropyl group. The same motif we find in Val at the bottom of the first column, as also, *mutatis mutandis*, a motif of isopropyl and isobutyl group branching we find at the top of the last column (Cys, Trp).

With this movement – downward, upward, lateral and diagonal, we have the resolution for the position of the serine, the explanation of how it is located in two distant and chemically different places in the GCT. Serine is at the beginning of the second column as the diagonal adjacent to leucine. On the other hand, serine is adjacent to arginine, and arginine is located at the second end of the Leu-Arg complementarity line ("1-14" in Tables 1 & 2). It becomes clear that in the same serine column there is cysteine, which is its chalcogenic analog (both elements, oxygen and sulfur are in VI group of PSE).

*

At the beginning, as well as at the end of GCT, we have relations between two and two AAs, with the minimum change: Phe-Tyr and Gly-Ala, respectively, as explained in the Scenario in Section 2. We have the complementarity of two and two AAs, but not through the addition operation, but through the subtraction on the Binary tree (15 - 7 = 8 and 8 - 0 = 8). Tyrosine as the derivative of Phenilalanine, like alanine as derivative of glycinne, both represent the minimal possible change. Hence, this is complementarity by *dissimilarity in similarity*.

Thus, one can notice that this complementarity expresses relationships between the first and third, as well as between the second and fourth columns in the GCT. The cases that follow, listed simultaneously, in the upper and lower parts of Table 1 can be analyzed in the same or similar manner, as to "chemical eyes" are directly apparent. [But this remains for some other occasions if it appears that there is of an interest of the scientific public.]

The specificity of the chemical complementarity, which we are talking about here, is expressed as complementarity of the nucleotide doublets as well as the corresponding amino acids on each two vertices on the Boolean hypercube ($B^4$) whose binary values give the sum 1111 in the binary record, corresponding to the number 15 in decimal one (Table 2).

**5. The polarity in relation to the cipher key.** In previous sections, arguments are given for making the cipher of the genetic code a specific chemism in the form of a specific chemical complementarity; and that the cipher's key represents a sequence of codon positions and correspondent amino acids on the binary-code tree of the genetic code (Rakočević, 1988, Fig. 31 on page 120; 1998, Fig. 1 on page 284). The following four Tables (3.1, 3.2 & 4.1, 4.2) contain the result, the final result, which supports that evidence. What is common in these tables is that it is directly apparent that the key of cipher splits AAs into two groups: a group in which the AAs are strictly differentiated into polar and nonpolar, and another group, in which this is not the case.

In the dark spaces of Table 3.1, the separation into polar and nonpolar AAs is complete. On the other hand, in the internal unshielded spaces, the separation is such that on one side we have a little polar AAs, and on the other side, the very polar. There is only one case where one opposite to the other we have non-polar AAs: a highly non-polar valine versus far less polar cysteine; and to make the separation more complete, tryptophan is added.

The eight amino acids, which correspond to eight "one-meaning" nucleotide doublets (AAs in the upper part of Table 3.1), make sense to name – the single amino acids. In such a case, tyrosine joins them because it is also the only AA in the four-codon UA space (Table 3.2).[10] As we see, this small change, this "tongue on the scale" causes new sophisticated shading and balancing, making "jumps" from one system to another, from one arrangement to another.

The changing of Table 3.1 in Table 4.1 is that purine encoding AAs are excluded, those that are encoded by codons that have purine in the third position. The result (number of atoms) in the upper part of the two Tables has not changed since the same AAs remained. But in the lower part it is. Instead of 88, now we have 44 (which is again a symmetry and balance!). On the other hand, instead of 85, we now have 31, correspondent with the result 31 in the upper part. The difference 85 - 31 = 54 is in correspondence with the result of 44 atoms. The whole thing becomes clear when we compare Table 3.1 with Table 4.2 where pyrimidine encoding AAs are excluded: we have a balance in one crossing: 54 vs 44 and 41 vs 31.

The differences from the aspect of the inclusion / exclusion of the block of Py / Pu amino acids have also been shown to be significant in other arrangements of the genetic code. So, this was demonstrated in the case of the classification of AAs into two classes, handled by two classes of enzymes of aminoacyl-tRNA synthetases (Damjanović and Rakočević, 2005, 2006, 2007), and also in the case of the analysis of "p-adic model of ... genetic code " (Dragovich et al., 2006, 2010, 2017).

*

In Table 3.1, it should be noted that at the main "block" of the change, at positions "1-14", there are amino acids L, S, R, precisely those that are doubled in the set of "23" AAs. Only with their duplication is it easier to nuance and balance chemical properties, which could be a kind of "intelligent design" (Box 2). Now, the meaning of their duplication is seen: non-polar Leu vs polar Ser & Arg. In sequence "0-15", we again have Le to "help" in establishing a relationship between non-polar Phe and polar Gly. The same applies to Arg, which, as polar, reappears in sequence "4-11" versus non-polar Ile and Met;

---

[10] In this definition, it is only important that the doublet UA can encode only one AA, irrespective of the fact that it can encode also the termination signal.

Finally, Ser is also in the sequence of positions "2-13" in order to be, as slightly polarized against the high polar Asp and Glu.

In Tables 3.1 & 3.2 and 4.1 & 6.2, we also showed the nuancing and balancing in the number of atoms in the distinction of AAs to "single AAs" and "double AAs". These results are in relation to results we showed in a previous work (Rakočević, 2004b, Tables 7 & 8) where the nuancing and balancing within the genetic code appear through seven different parameters.

**Box 2.** *The Spontaneous Intelligent Design*

Castro-Chavez, 2010, p. 718: "We can conclude that the genetic code is an intelligent design that maximizes variation while minimizing harmful mutations."

Rakočević, 2013 and 2015, p. 18: “With insight into the results … one is forced to propose a hypothesis (for further researches) that here, there really is a kind of intelligent design; not the original intelligent design, dealing with the question – intelligent design or evolution (Pullen, 2005), which is rightly criticized by F.S. Collins (2006). Here, there could be such an intelligent design, which we could call “Spontaneous Intelligent Design” (SPID) that is consistent with that design which was presented by F. Castro-Chavez (2010), and is also in accordance with the Darwinism. ... Actually, it can be expected that the hypothetical SPID, contained in the results …, is in accordance with an identical (or similar?) SPID, presented in the only diagram, in Darwin's book “Origin of Species” (Darwin, 1859), as we have shown through an analysis of that diagram in one of our books (Rakočević, 1994; www.rakocevcode.rs). [In the case of the statement that spontaneity and intelligent design are mutually opposite, one must ask the question: isn’t it true that human intelligence is the result of a spontaneous evolutionary process?]”

The divisions, given by one and the same hatching within the four Tables (3.1 & 3.2 and 6.1 & 6.2) show that over the same division there is a specific proof of the correctness of the hypothesis about the necessity of distinguishing the genetic code on the code, the cipher and the key of the cipher. In a certain way, one confirms the other: the values of the hydropathy index confirm the validity of the cipher key, and the cipher key confirms the validity of the experimentally determined values of the hydropathy index (Kyte & Doolitle, 1982).

**6. concluding remarks.** The facts about a specific chemical complementarity of the constituents of the genetic code, given throughout this paper provide evidence to support the hypothesis, given in the title of this paper that the genetic code can be interpreted as the unity of the three entities: the code, the cipher of the code and the key of the cipher. Just only with this understanding, we can find, within the two main Genetic Code Tables (of the nucleotide doublets and nucleotide Triplets) the sophisticated nuancing and balancing in the properties of the constituents of GC, including the balance of the number of molecules, atoms, and nucleons.

All this also confirms our hypothesis, given in one of the previous paper (Rakočević, 2004b,) that the genetic code, from the beginning, in prebiotic conditions, was complete. Out of the millions of possible aggregations of molecules, the potential builders of GC, on "the card of life" played the one that potentially possessed all the chemical complementarities, we have exposed here.[11]

Our expectation is that this work could be joined to works that are on the way to solving open problems of existence and the essence of genetic code; such problems related to the search for answers to questions of origin and evolution of life; in particular, those works that open up new fronts of research in biology, extending the theme of genetic code on topics of *The codes of life*, or, more broadly, to *The biological codes* (Barbieri, Hofmeyr et al., 2008, 2018).

**Acknowledgements**

I am grateful to Vladimir Ajdačić, Branko Dragovich, Djuro Koruga, Nataša Mišić, Tidjani Négadi and Dejan Raković for helpful, stimulating discussions and benevolent critique.

[11] "Each of that aggregations could (and must) have its own ‘evolution’, but only one could have been selected – the one that gained the characteristic of self-reproduction (by which, through trial, error and success it became everything); all other, not selected, could not have any chance ..., they became nothing" (Rakočević, 2004b, p. 232).

## TABLES

**Table 1.** The Canonical Invariant Sistem (CIS) of codons and corresponding amino acids, according their positions on the binary-code tree (Rakočević, 1998)

| | | | | | | | | |
|---|---|---|---|---|---|---|---|---|
| **0** | UUN | F, L | **8** | UAN | Y | **15** | GGN | G |
| **1** | CUN | L | **9** | CAN | H, Q | **14** | AGN | S, R |
| **2** | UCN | S | **10** | UGN | C, W | **13** | GAN | D, E |
| **3** | CCN | P | **11** | CGN | R | **12** | AAN | N, K |
| | | 53<br>46 | | | 77<br>63 | | | 63<br>77 |
| **4** | AUN | I, M | **12** | AAN | N, K | **11** | CGN | R |
| **5** | GUN | V | **13** | GAN | D, E | **10** | UGN | C, W |
| **6** | ACN | T | **14** | AGN | S, R | **9** | CAN | H, Q |
| **7** | GCN | A | **15** | GGN | G | **8** | UAN | Y |
| (53+46 = 119 - 20) (53+77 = 120+10)<br>(77+63 = 120 +20) (46+63 = 119 - 10) | | | | | | Left 46 56 Right<br>53 84 | | |
| (53+63 =115+1) (46 + 77 = 124-1) | | | | | | 46 + 56 = 102<br>53 + 84 = 102 + 35 | | |

The number of atoms in the upper and lower part of the Table corresponds to the number of atoms in GCT. Diagonal result [115+1 & 124-1 vs 115 & 124 in (Rakočević, 2017c, Chapter VI, Table 1)]. A "surplus" of 35 atoms is a "balance fraction" which, when passing from (102 + 102 = 204 atoms) in a set of 20 AAs to 239 atoms in a set of 23 AAs, corresponds to the quantity for three doubled AAs (L13 + S05 + R17 = 35). In relation to original Rumer's Table where within amino acids (their side chains) there are: up/down: 119/120 atoms, and here there is a change for ±10.

**Table 2.** The vertical CIS display into one-meaning and two-meaning nucleotide doublets and corresponding amino acids

| an | $on_2$ | $on_1$ | $c_1$ | aa | s | aa | $c_2$ | $on_1$ | $on_2$ | an |
|---|---|---|---|---|---|---|---|---|---|---|
| 13 | **1** | 0001 | CUN | L | (15) 1111 | S, R | AGN | 1110 | **14** | 22 |
| 05 | **2** | 0010 | UCN | S | (15) 1111 | D, E | GAN | 1101 | **13** | 17 |
| 08 | **3** | 0011 | CCN | P | (15) 1111 | N, K | AAN | 1100 | **12** | 23 |
| 10 | **5** | 0101 | GUN | V | (15) 1111 | C, W | UGN | 1010 | **10** | 23 |
| 36 | | | | | | | | | | 85 |
| 08 | **6** | 0110 | ACN | T | (15) 1111 | H, Q | CAN | 1001 | **9** | 22 |
| 04 | **7** | 0111 | GCN | A | (15) 1111 | Y, ct | UAN | 1000 | **8** | 15 |
| 17 | **11** | 1011 | CGN | R | (15) 1111 | I, M | AUN | 0100 | **4** | 24 |
| 01 | **15** | 1111 | GGN | G | (15) 1111 | F, L | UUN | 0000 | **0** | 27 |
| 30 | | | | | | | | | | 88 |
| | (333 + 592 = 925) (1110 +1110)<br>259 000 | | | | | | | | | |
| | (36 + 88 = **124**) (30 +85 = **115**) | | | | | | | | | |

The designations: an – number of atoms within amino acid side chain; $on_1$ – ordinal number in binary records; $on_2$ – ordinal number in decimal records; $c_1$ – codons with containing one-meaning nucleotide doublets; $c_2$ – codons with containing two-meaning nucleotide doublets; aa – amino acids. In relation to original Rumer's Table where within amino acids (their side chains) there are: left/right fiagonally: 114/125 atoms, and here there is a change for ±10.

**Table 3.1.** The horizontal CIS display into one-meaning and two-meaning nucleotide doublets and corresponding amino acids (I)

| 1 | 2 | 3 | 5 | 6 | 7 | 11 | 15 |
|---|---|---|---|---|---|---|---|
| L<br>+3.8 | S<br>-0.8 | P<br>-1.6 | V<br>+4.2 | T<br>-0.7 | A<br>+1.8 | R<br>-4.5 | G<br>-0.4 |
| -0.8 -4.5<br>S R | -3.5 -3.5<br>D E | -3.5 -3.9<br>N K | +2.5 -0.9<br>C W | -3.2 -3.5<br>H Q | -1.3<br>Y | +4.5 +1.9<br>I M | +2.8 +3.8<br>F L |
| 14 | 13 | 12 | 10 | 9 | 8 | 4 | 0 |

| | | | |
|---|---|---|---|
| | LARG = 36-1<br>SRYIMFL = 88 | 35+85 = 120<br>88+31 = 119 | SPVT = 30+1<br>DENKCWHQ = 85 |

All is the same as in Table 2, with an additional distinction of amino acid molecules through their polarities, measured by the hydropathy index (Kyte & Doolittle, 1982). Notice a diagonal balance in which it is shown that number of atoms in amino acid singlets and doublets is determined by the middle pair of the total number of atoms within the set of 23 AAs (0, 1, 2 , 3, 4, ..., **119-120**, ..., 235, 236, 237, 238, 239); all that in a strict distinction into polar and non-polar AAs.

**Table 3.2.** The horizontal CIS display into one-meaning and two-meaning nucleotide doublets and corresponding amino acids (II)

| 1 | 2 | 3 | 5 | 6 | 7 | 8 | 11 | 15 |
|---|---|---|---|---|---|---|---|---|
| L | S | P | V | T | A | Y | R | G |
| +3.8 | -0.8 | -1.6 | +4.2 | -0.7 | +1.8 | -1.3 | -4.5 | -0.4 |
| -0.8 -4.5 | -3.5 -3.5 | -3.5 -3.9 | +2.5 -0.9 | -3.2 -3.5 | | ±0.0 | +4.5 +1.9 | +2.8 +3.8 |
| S R | D E | N K | C W | H Q | | Θ | I M | F L |
| 14 | 13 | 12 | 10 | 9 | | 8 | 4 | 0 |

| | | | |
|---|---|---|---|
| F.R. | LAYRG = 50 | 50 + 31 = 81 | SPVT = 31 |
| II | SRIMFL = 73 | 73+85 = 123 + 35 | DENKCWHQ = 85 |

All is the same as in Table 3.1, with an exception: the amino acid tyrosine is above instead down. Notice here a specific balance in which it is shown that the number of atoms in amino acid singlets (9 AAs above) equals 81 as in 10 AAs of class II, handled by class II of enzymes of aminoacyl-tRNA synthetases; On the other hand, the quantity of the number of atoms in 14 doublet AAs is 123 + 35, where the number 123 corresponds to the number of atoms in class I of AAs, handled by class I of enzymes aminoacyl-tRNA synthetases and the number 35 is a "surplus" of 35 atoms is a "balance fraction" which, when passing from 204 atoms in the set of 20 AAs to 239 atoms in the set of 23 AAs, corresponds to the quantity for three doubled AAs (L13 + S05 + R17 = 35). [About classification into two classes of AAs, handled by two classes of enzymes aminoacyl-tRNA synthetases, one can see in: Wetzel, 1995; Rakočević, 1997a; Rakočević, 1998, Survey 4, p. 290.]

**Table 4.1.** The horizontal CIS display into one-meaning and two-meaning nucleotide doublets and corresponding amino acids (III)

| | | | | | | | |
|---|---|---|---|---|---|---|---|
| L | S | P | V | T | A | R | G |
| +2.8 | -0.8 | -1.6 | +4.2 | -0.7 | +1.8 | -4.5 | -0.4 |
| -0.8 | -3.5 | -3.5 | +2.5 | -3.2 | -1.3 | +4.5 | +2.8 |
| S | D | N | C | H | Y | I | F |

| | | | |
|---|---|---|---|
| F.R.<br>III | LARG = 35<br>SYIF = **44** | SPVT = 31<br>DNCH = **31** | |

It is all the same as in Table 3.1, except that purine-coding amino acids are excluded in the bottom row.

**Table 4.2.** The horizontal CIS display into one-meaning and two-meaning nucleotide doublets and corresponding amino acids (IV)

| | | | | | | | |
|---|---|---|---|---|---|---|---|
| L | S | P | V | T | A | R | G |
| +2.8 | -0.8 | -1.6 | +4.2 | -0.7 | +1.8 | -4.5 | -0.4 |
| -4.5 | -3.5 | -3.9 | -0.9 | -3.5 | ±0.0 | +1.9 | +3.8 |
| R | E | K | W | Q | Θ | M | L |

| | | | |
|---|---|---|---|
| F.R.<br>IV | LARG = 35<br>RML = **41** | SPVT = 31<br>EKWQ = **54** | |
| (44 + 10 = 54) (31 + 10 = 41) | | | |

It is all the same as in Table 3.1, except that pyrimidine-coding amino acids are excluded in the bottom row. In comparison Table 4.1 and 4.2 notice the balance in the atom number: 31 & 44 versus 41 & 54.